\documentclass[pra,twocolumn,a4paper,groupedaddress,showpacs,floatfix,aps,10pt,
longbibliography,nobibnotes]{revtex4-1}
\usepackage{graphicx,amsmath,amssymb,amsfonts,dsfont,subfigure,color}

\usepackage{cleveref}

\newcommand{\mc}[1]{\ensuremath{\mathcal{#1}}}

\begin{document}

\title{Microwave-assisted Rydberg electromagnetically induced transparency}

\author{Thibault Vogt${}^{1,2}$}
\author{Christian Gross${}^{1}$}
\author{T. F. Gallagher${}^{3}$}
\author{Wenhui Li${}^{1,4}$}

\affiliation{Centre for Quantum Technologies, National University of Singapore, 3 Science Drive 2, Singapore 117543${}^1$}
\affiliation{MajuLab, CNRS-UNS-NUS-NTU International Joint Research Unit UMI 3654, Singapore 117543${}^2$}
\affiliation{Department of Physics, University of Virginia, Charlottesville, Virginia 22904, USA${}^3$}
\affiliation{Department of Physics, National University of Singapore, 117542, Singapore${}^4$}




\begin{abstract}
We demonstrate electromagnetically induced transparency (EIT) in a four-level cascade-like system, where the two upper levels are Rydberg states coupled by a microwave field. A two-photon transition consisting of an off-resonant microwave field and an off-resonant optical field forms an effective coupling field to induce transparency of the probe light. We characterize the Rabi frequency of the effective coupling field, as well as the EIT microwave spectra. The results show that microwave assisted EIT allows us to efficiently access Rydberg states with relatively high orbital angular momentum $\ell=3$, which is promising for the study of exotic Rydberg molecular states.

\end{abstract}


\maketitle

\smallskip
%

Electromagnetically induced transparency involving Rydberg states (Rydberg EIT) has opened up new research avenues for Rydberg physics and for nonlinear optics~\cite{pritchard2010cooperative,firstenberg2016nonlinear}. Rydberg EIT is an all-optical method to access the exaggerated properties of Rydberg atoms, and very well suited to probing the effects of long-range interactions between Rydberg atoms. The extreme sensitivity of Rydberg EIT to these interactions manifests itself in fascinating phenomena such as photon blockade and effective photon-photon interactions, which have direct applications for deterministic single-photon sources, photonic quantum gates, and imaging of Rydberg atoms~\cite{dudin2012strongly,peyronel2012quantum,firstenberg2013attractive,tiarks2016optical,busche2017contactless,gunter2013observing}. Furthermore, the simplicity of Rydberg EIT makes it an excellent platform for all-optical spectroscopy, which often requires only thermal vapors as atomic sample. 
 It has been used successfully for measurements of atomic and molecular transitions, in electrometry of static or microwave fields, and for polarization measurements of microwave fields ~\cite{grimmel2015measurement,mirgorodskiy2017electromagnetically,PhysRevLett.112.026101,PhysRevLett.111.063001,sedlacek2012microwave,simons2016using}.

Electromagnetically induced transparency (EIT) usually involves two electromagnetic fields, probe and coupling~\cite{fleischhauer:05}. In Rydberg EIT, the probe and coupling fields couple a ground state to a long-lived Rydberg state via an intermediate short-lived level in a ladder excitation scheme~\cite{weatherill:08,mack2011measurement}.
For this reason, Rydberg EIT relies on $nS$ or $nD$ Rydberg states (with the ground state in an $S$ state), where $n$ is the principal quantum number.
EIT dark state dressing involving $nP$ Rydberg states has also been investigated experimentally, where a third resonant field (microwave or optical) dresses the EIT dark state and results in a split transparency resonance for the probe field~\cite{sedlacek2012microwave,simons2016using,tanasittikosol2011microwave}.

In this letter we demonstrate three-photon EIT involving $nP$ or $nF$ Rydberg states, where two fields, microwave and optical, are used in a two-photon resonance configuration via an off-resonant Rydberg state $n'D$, and form an effective coupling field for the probe light to propagate without absorption. We achieve large effective coupling Rabi frequencies thanks to the huge electric dipole matrix elements of the transitions between neighboring Rydberg states~\cite{gallagher:ryd}. 
This report and a related one in Ref.~\cite{tate2018microwave} represent the first use of the large dipole moments for microwave-optical excitation of Rydberg states. The transparencies achieved in this experiment are similar to that obtained in three-level Rydberg EIT. This work opens up new opportunities for spectroscopic measurement. For example, with direct excitation of the Rb $nF$ states, it should be possible to optically detect trilobite Rb$_2$ molecules. In these dimers, one of the two atoms is in a superposition of the Rb $nF$ state and the higher angular momentum states of the same $n$~\cite{greene:00}. Furthermore, this work represents important progress towards using three-photon Rydberg EIT in coherent microwave-optical frequency conversion~\cite{kiffner2016two,han2017free}.

\smallskip

\begin{figure}[htbp]
\centering
\includegraphics[width=\linewidth]{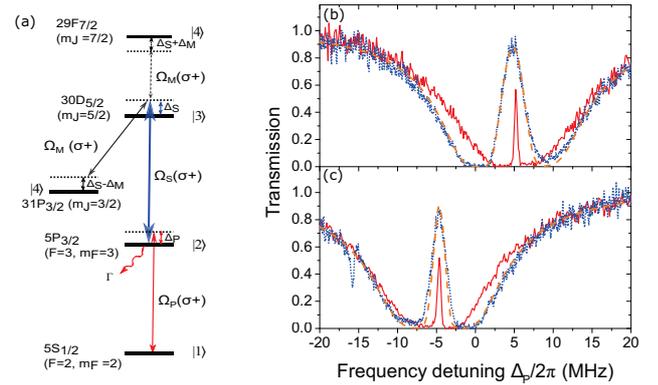}
\caption{(a) Relevant energy levels of $^{87}$Rb and dipole coupled transitions. The microwave transition in the EIT pathway to the $31P_{3/2}$ state is indicated by a black solid arrow, and by a dashed arrow for EIT to the $29F_{7/2}$ state. (b) and (c) EIT spectra versus $\Delta_P$ for the $31P_{3/2}$ and $29F_{7/2}$ states, respectively. In (b), the microwave field Rabi frequencies of the different spectra are $\Omega_M/ 2 \pi=6.7$~MHz (solid line), and $24.8$~MHz (blue dotted line). In (c), the microwave field Rabi frequencies are $\Omega_M/ 2 \pi=6.7$~MHz (solid line), and $16.7$~MHz (blue dotted line). The dashed lines are simulated spectra from Maxwell-Bloch's equations based on four-energy-level atoms.}
\label{Fig1}
\end{figure}

The excitation scheme is shown in Fig.~\ref{Fig1}(a). A weak probe field P, a relatively strong optical field S, and a microwave field M couple with three consecutive electric dipole transitions of $^{87}$Rb atoms, $|1\rangle \rightarrow |2\rangle$, $|2\rangle \rightarrow |3\rangle$, and $|3\rangle \rightarrow |4\rangle$, where $|1\rangle \equiv |5S_{1/2},F=2,m_F=2\rangle$, $|2\rangle \equiv |5P_{3/2},F=3,m_F=3\rangle$, $|3\rangle \equiv |30D_{5/2},m_J=5/2\rangle$, and state $|4\rangle$ stands either for $|31P_{3/2}, m_J=3/2\rangle$, or for $|29F_{7/2}, m_J=7/2 \rangle$. The wavelengths of the P and S fields are $\lambda_P \sim 780\ \textrm{nm}$ and $\lambda_S \sim 482\ \textrm{nm}$, respectively, whereas the microwave frequency is $\sim84\ \textrm{GHz}$ for the excitation to $|31P_{3/2}, m_J=3/2\rangle$, and $\sim90.5\ \textrm{GHz}$ for the excitation to $|29F_{7/2}, m_J=7/2 \rangle$.
The Rabi frequency of the field S is kept constant in this experiment, with a value of $\Omega_S/ 2 \pi  = 30\pm2$~MHz.  We blue-detune the field S for exciting the $|31P_{3/2}, m_J=3/2\rangle$ state ($\Delta_S=32\ \textrm{MHz}$) and red-detune it for $|29F_{7/2}, m_J=7/2 \rangle$ ($\Delta_S=-32\ \textrm{MHz}$). The detuning of the microwave field M is positive in both cases, as it is adjusted to drive the two-photon transition $|2\rangle \rightarrow |4\rangle$ resonantly. 

The experimental apparatus and various methods for parameter calibrations have been described in Refs. \cite{han2015lensing,han2017free}. In short, a Gaussian-distributed atomic cloud, with temperature $T\sim 50\ \mu\textrm{K}$, $1/e^2$ radius $w_z \sim 1.2\ \textrm{mm}$, and peak atomic density $n_{at}^{(0)}=1.9 \times 10^{10}\ \textrm{cm}^{-3}$, is prepared in state $|1\rangle$ by optical pumping. The cloud is then illuminated for 1~ms with the microwave and optical fields. The P and S Gaussian fields of $\sigma^+$ polarization counterpropagate along the quantization axis $z$, and are focused on the atomic cloud with beam waists of $25\ \mu\textrm{m}$ and $54\ \mu\textrm{m}$, respectively. The microwave field is emitted from the side, perpendicularly to $z$, and is of equal superposition of $\sigma^+$ and $\sigma^-$ polarizations. However, the presence of an applied magnetic field of 6.1~G along $z$ to split the Zeeman sublevels, and the choice of a positive detuning $\Delta_M>0$, ensure that the weak transitions which couple to the $\sigma^-$ microwave field component are sufficiently off-resonance~\cite{Note1}.
 The spectra of the P field power $P_P$ transmitted through the atomic cloud are recorded with a photomultiplier tube~(Hamamatsu R636-10). They are obtained by continuously scanning either the P field detuning $\Delta_P$ or the M field detuning $\Delta_M$ during the 1~ms time window. The P field transmission spectra are then plotted as $P_P/P_{P}^{(0)}$, where $P_{P}^{(0)}=0.13\ \textrm{nW}$ is the input P field power.

\smallskip

A selection of transmission spectra versus $\Delta_P$ are shown in Fig.~\ref{Fig1}(b), for which $|4\rangle \equiv |31P_{3/2},m_J=3/2\rangle$. The EIT peak is positioned at the center of the broad absorption profile, caused by spontaneous decay from state $|2\rangle$ with rate $\Gamma= 2\pi \times 6.07\ \textrm{MHz}$. 
The detuned S and M fields induce AC Stark shifts for the levels $|2\rangle$ and $|4\rangle$, shifting both the central position of the absorption profile and the microwave frequency at which EIT occurs. In condition of near two-photon resonance, those two shifts are $\delta_2\approx \Omega_S^2/ 4 \Delta_S$ and $\delta_4\approx\Omega_M^2/ 4 \Delta_S $, respectively, and the EIT is obtained when $\Delta_P+\Delta_S-\Delta_M -\delta_4\approx 0$, as shown further below. Thus, in each of the spectra displayed in Fig.~\ref{Fig1}(b), the detuning $\Delta_M$ is slightly readjusted to position the transparency peak at the center of the absorption profile. Similar results are obtained when $|4\rangle \equiv |29F_{7/2},m_J=7/2\rangle$ (see Fig.~\ref{Fig1}(c)). In this case, the AC Stark shift for the energy level $|2\rangle$ is opposite since $\Delta_S$ is of opposite sign.
In both cases, the EIT spectra are very similar to those obtained in a three-level ladder-type configuration. This is indeed expected if the population in state $|3\rangle$ is negligible, which means the S and M fields form an effective coupling field that opens a transparency window for the probe light to propagate through the atomic cloud without absorption.

In order to give a more quantitative interpretation of our results, we calculate the transmission of the probe field $\mc{E}_{P}\left( \vec r\right)$ with Maxwell's equation.
After applying the slow envelope approximation and neglecting lensing effects or diffraction~\cite{han2015lensing}, Maxwell's equation simplifies to

\begin{equation}
\partial_{z} \mc{E}_{P}\left( \vec r\right)=   i \frac {\pi } {\lambda_P} \chi \left( \vec r \right) \mc{E}_{P}\left( \vec r \right).
\label{Maxwell}
\end{equation}
Here $\chi \left( \vec r \right)$ is the atomic susceptibility at position $\vec r =(x,y,z)$ in the atomic sample. Since the experiment is performed with a weak probe intensity,  of input peak Rabi frequency $\Omega_P^{(0)} \sim  2 \pi \times 0.4\ \textrm{MHz} \ll \Gamma$, $\chi$ may be simply taken at the first order in $\Omega_P$. 
The susceptibility for a four-level system then writes at first order~\cite{sandhya1997atomic}
\begin{equation}	
\chi_{4l} =i \frac{n_{at} \Gamma \sigma_0 \lambda_P }{ 4 \pi \left(\gamma_{12}-i \left(\Delta _P -\frac{\Omega_S^2}{4 \Delta_S L_{13}}\right)+\frac{\Omega_S^2 \Omega_M^2}{16 \Delta_S^2 L_{13} \ \left(L_{13}\, L_{14} + i \, \delta_4 \right)}\right)} \label{susceptibility4l},
\end{equation}
where $n_{at}$ is the atomic density, $\sigma_0=3 \lambda_P^2/ 2 \pi$ is the resonant scattering cross-section, $\gamma_{12} \sim \Gamma/2$, and $L_{13}$ and $L_{14}$ are coefficients that depend on the frequency detunings and some dephasing rates. Namely, we have $L_{13}=1+\left( \Delta_P + i \gamma_{13} \right)/\Delta_S$ and $L_{14}=\gamma_{14} -i \left(\Delta_P+\Delta_S+\epsilon \Delta_M \right)$, where $\gamma_{ij}$ is the dephasing rate of the atomic coherence between states $|i\rangle$ and $|j\rangle$, and where $\epsilon=-1$ for the $31P_{3/2}$ state and $\epsilon=+1$ for the $29F_{7/2}$ state. The main source of dephasing for $\gamma_{13}$ and $\gamma_{14}$ is laser phase noise, hence they will be considered as constant $\sim 2 \pi \times 100\ \textrm{kHz}$. The dashed lines in all the figures represent theoretical curves calculated with \cref{Maxwell,susceptibility4l}. 
All the parameters are taken from experimental measurements, while a slight adjustment of the detunings is still permitted, within experimental error bars ($\pm 0.5\ \textrm{MHz}$ for the optical ones and $\pm 0.2\ \textrm{MHz}$ for the microwave one). The good agreement with the experiment demonstrates that our independent four-level atom model captures the main physics of the system. Resonant dipole-dipole interactions that could occur due to residual occupation of state $|3\rangle$ are thus negligible. Moreover, unwanted $\sigma^-$ transitions have little effect, although one of them does appear in the $29F_{7/2}$ spectra in Fig.~\ref{Fig1}(c) as a small side dip.
The experimental spectra are slightly asymmetric, which is not predicted by our model. 
We have verified theoretically that optical beam imperfections or a slight misalignment of the counterpropagating beams create an inhomogeneous Stark shift of level $|2\rangle$, which would be consistent with the observed asymmetry.

\begin{figure}[htbp]
\centering
\includegraphics[width=\linewidth]{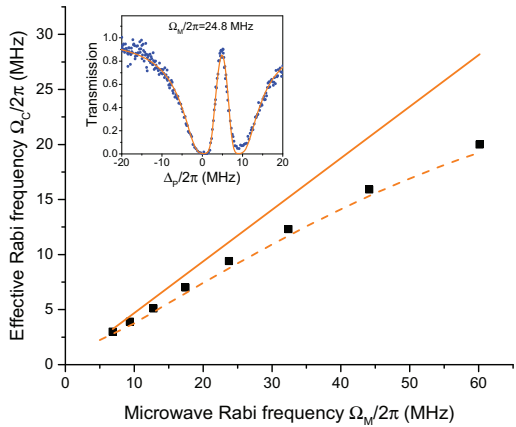}
\caption{Effective coupling Rabi frequency as a function of input microwave Rabi frequency, retrieved from spectra acquired as in Fig.~\ref{Fig1}~(b). The effective Rabi frequencies (squares) are obtained from fits of \cref{Maxwell,susceptibility3l} to the data, as shown in the inset for the particular case $\Omega_M/ 2 \pi=24.8\ \textrm{MHz}$. The dashed line shows the effective coupling Rabi frequencies extracted from spectra simulated with the four-level model and fitted by \eqref{susceptibility3l}, while the solid line corresponds to the approximation $\Omega_C = \Omega_S \Omega_M/ 2 \Delta_S$.}
\label{Fig2}
\end{figure}

In the limit of large detuning $\Delta_S$ compared to $\Delta_P$, $\Delta_S + \epsilon \Delta_M$, $\gamma_{ij}$, $\delta_2$, and $\delta_4$, \eqref{susceptibility4l} reduces in a first approximation~\cite{Note2} to
\begin{equation}	
\chi_{4l} \approx i \frac{n_{at} \Gamma \sigma_0 \lambda_P}{ 4 \pi \left(\gamma_{12}-i \left(\Delta _P-\delta_2\right)+\frac{\Omega _C  ^2}{4 (\gamma _{14}-i (\Delta_P+\Delta _C-\delta_4))}\right)} \label{susceptibility3l},
\end{equation}
where $\Omega_C = \Omega_S \Omega_M/ 2 \Delta_S$ is the Rabi frequency of the effective coupling field, and $\Delta_C=\Delta _S+ \epsilon \Delta_M$ its detuning, as also predicted by the adiabatic elimination approximation usually applied to describe multi-photon excitations~\cite{Paulisch2014}. \eqref{susceptibility3l} corresponds effectively to the susceptibility of a three-level system, where, however, the resonances for the P and C fields are shifted by the AC Stark effect~\cite{fleischhauer:05}. 
We use \cref{Maxwell,susceptibility3l} with $\delta_2$, $\delta_4$, $\Omega_C$, $\gamma_{14}$, and $n_{at}^{(0)}$ as adjustable parameters to fit the spectra in Fig.~\ref{Fig1}(b) and estimate the effective coupling Rabi frequencies $\Omega_C$ that we achieve with this setup (Fig. \ref{Fig2}). The fits are very good (inset of Fig.~\ref{Fig2}), and large $\Omega_C$ are measured as a function of input microwave Rabi frequency $\Omega_M$, up to $\Omega_C/2 \pi \approx 20\ \textrm{MHz}$. It appears that there is a significant departure for large $\Omega_M$ from the simple estimate $\Omega_C \sim \Omega_S \Omega_M/ 2 \Delta_S$ (solid line in Fig.~\ref{Fig2}). This comes as no surprise when $\Omega_M > \Delta_S$, in which condition some of the approximations used to derive \eqref{susceptibility3l} are no longer valid.

\begin{figure}[htbp]
\centering
\includegraphics[width=\linewidth]{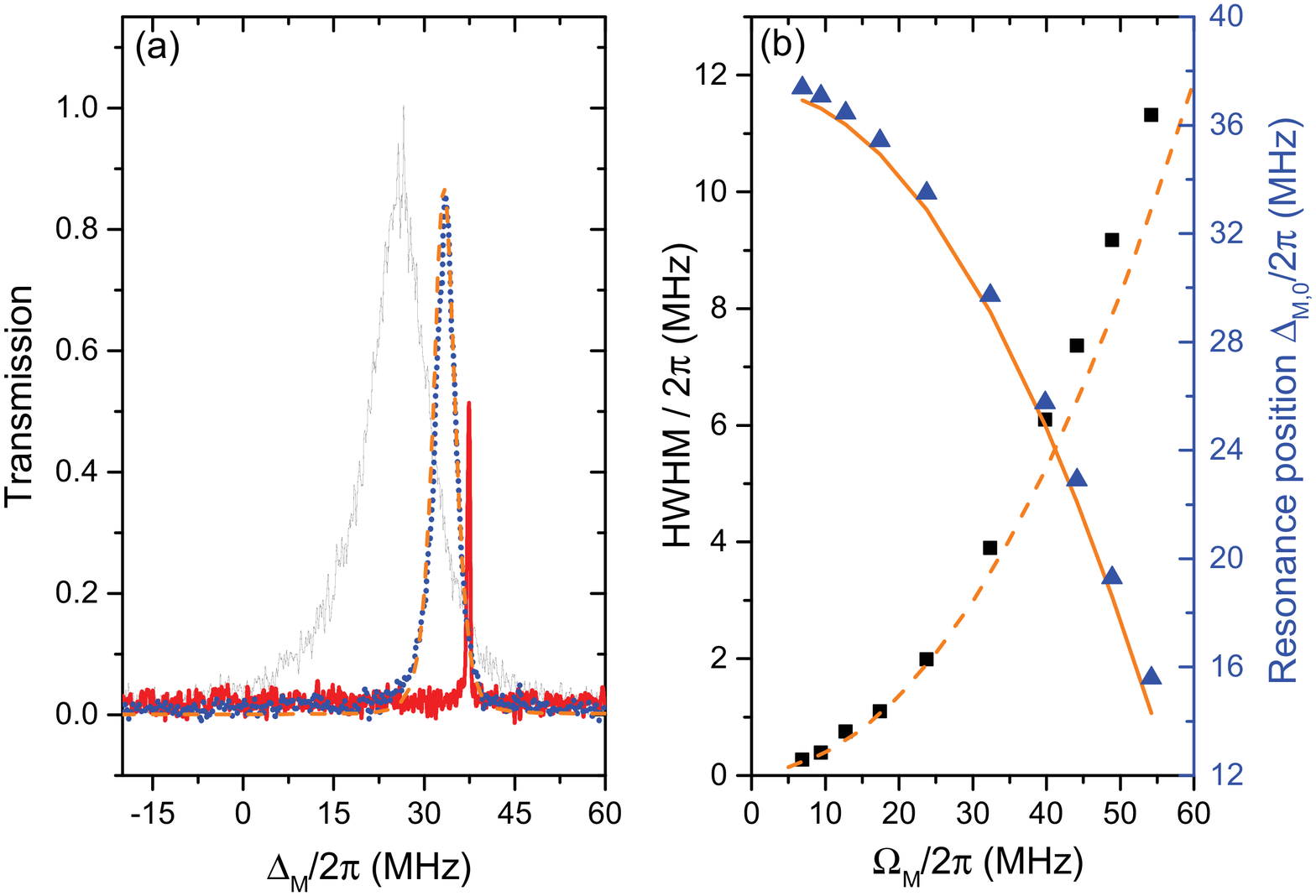}
\caption{ (a) EIT versus microwave field detuning. Typical spectra obtained for microwave field Rabi frequencies $\Omega_M/2 \pi=6.7$~MHz (thick red solid line), 24.8~MHz (blue dotted line), and 39.8~MHz (thin black solid line). The dashed line is the result of Maxwell-Bloch's equations based on the four-level model, for the case $\Omega_M/2 \pi=24.8$~MHz. (b) Half width at half maximum (HWHM) extracted from experimental spectra (squares), and microwave field detuning $\Delta_{M,0}$ at the center of the resonance (triangles). The dashed line is the result of the four-level model, while the solid line corresponds to the quadratic approximation of the AC Stark shift (see text).}
\label{Fig3}
\end{figure}

Next, we analyze EIT using a different approach, by changing $\Delta_M$ instead of $\Delta_P$. The advantages of using microwave fields are indeed the simplicity and the stability of the available synthesizers, with the possibility to scan frequencies over very wide ranges. In Fig.~\ref{Fig3}(a), we display a few microwave spectra, where $\Delta_M$ is scanned across the EIT resonance, while we maintain $\Delta_P / 2\pi =5.3\ \textrm{MHz}$ in order to compensate for the AC Stark shift. The spectra are symmetric, almost squared Lorentzian in shape, and agree well with the solution of \cref{Maxwell,susceptibility4l}. The frequency shifts and widths of the resonances are recorded in Fig.~\ref{Fig3}(b). The frequency shifts show a quadratic dependence versus $\Omega_M$, as expected from the simple estimate, $\Delta_{M,0}=\Delta_0-\Omega_M^2/ 4 \Delta_S $, where $\Delta_0= \Delta_P+\Delta_S$. This result shows that the measured AC Stark shifts follow the prediction obtained in the derivation of \eqref{susceptibility3l}, even when the effective Rabi frequency does not. The widths of the resonances are also quadratic versus $\Omega_M$, which is expected as it is known that the EIT linewidths are proportional to $\Omega_C^2$~\cite{fleischhauer:05}. Note that very similar results, not shown here, are obtained for the $29F_{7/2}$ state, up to $\Omega_M=25\ \textrm{MHz}$. Contrary to the $31P_{3/2}$ case, the weak $\sigma^-$ transitions become non negligible for larger $\Omega_M$. We have observed that larger $\Omega_M$ can be used to yield efficient EIT with the $29F_{7/2}$ state when $\Delta_S>0$. However this choice of S-field detuning yields effective Rabi frequencies that slightly deviate from the ones expected with our simple four-level picture of the excitation, and would most likely require a theoretical description with more energy levels.

\smallskip

In summary, we have demonstrated that microwave fields can be used to efficiently induce effective three-level EIT involving $nP$ or $nF$ Rydberg states. The effective Rabi frequencies, AC Stark shifts, and microwave frequency bandwidths have been analyzed. 
This work has important implication for spectroscopy and could be useful for the detection of Rydberg molecular states  \cite{li2011homonuclear,booth2015production,yu2013microwave,PhysRevLett.117.083401}.
Furthermore, three-photon Rydberg EIT corresponds to the solution found in Ref.~\cite{kiffner2016two} for optimizing microwave-optical conversion based on frequency mixing to reach near-unit photon conversion efficiencies.

\section*{Acknowledgments}

The authors acknowledge the
support by the National Research Foundation, Prime
Ministers Office, Singapore and the Ministry of Education, Singapore under the
Research Centres of Excellence programme. This work is supported by Singapore
Ministry of Education Academic Research Fund Tier 2 (Grant No.
MOE2015-T2-1-085).




\end{document}